\documentclass[]{spie}

\usepackage{aas_macros} 
\usepackage[export]{adjustbox}
\usepackage{amsfonts}
\usepackage{amsmath}
\usepackage{amssymb}
\usepackage{booktabs}
\usepackage{graphicx}
\usepackage{siunitx}
\usepackage[colorlinks=true, allcolors=blue]{hyperref}

\title{Spectrally dispersed non-redundant aperture-masking interferometry in the thermal infrared with LBTI/ALES}

\author[a]{T. A. Stuber}
\author[b]{J. M. Stone}
\author[a]{J. W. Isbell}
\author[c, d, e]{T. De Prins}
\author[a]{J. P. Scott}
\author[a]{J. A. Eisner}
\author[a]{J.~Carlson}
\author[a, f]{S. Ertel}
\affil[a]{Department of Astronomy and Steward Observatory, The University of Arizona, 933 North Cherry Avenue, Tucson, AZ 85721, USA}
\affil[b]{Department of Physics and Astronomy, University of Wyoming, 1000 East University Avenue, Laramie, WY 82071, USA}
\affil[c]{Institute of Astronomy, KU Leuven, Celestijnenlaan 200D, 3001 Leuven, Belgium}
\affil[d]{School of Mathematical and Physical Sciences, Macquarie University, Balaclava Road, Sydney, NSW 2109, Australia}
\affil[e]{Astrophysics and Space Technologies Research Centre, Macquarie University, Balaclava Road, Sydney, NSW 2109, Australia}
\affil[f]{Large Binocular Telescope Observatory, The University of Arizona, 933 North Cherry Avenue, Tucson, AZ 85721, USA}

\authorinfo{
    Further author information:\\
    T. A. S. e-mail: tstuber@arizona.edu\\
    LBTI Principal Investigator e-mail: lbtipi@lbto.org
}

\begin{document}


\maketitle


\begin{abstract}
Non-redundant aperture-masking interferometry maximizes the angular resolution
of a telescope by turning it into an interferometric array using pupil plane
masks. If the interference pattern on the detector is spectrally dispersed,
deeper contrasts of the astrophysical scenery and an improved coverage of its
spatial Fourier modes is achieved.\\
We combined a non-redundant aperture mask with the integral field spectrograph
ALES of LMIRcam at the Large Binocular Telescope Interferometer (LBTI). With
this setup, we recorded spectrally dispersed non-redundant aperture-masking
observations in the $L$ band ($\sim$\num{3}–\qty{4}{\um}).\\
We present commissioning data with which we recovered a known binary star and
derived contrast detection limits for further point-source components of
${\sim} \num{5e-3}$ for separations of ${\gtrsim} \qty{80}{mas}$. This is the
first time that an integral field spectrograph was combined with a
non-redundant aperture mask in the $L$ band. Opening up the thermal infrared
for this technique, this new capability enables the LBTI to efficiently survey
nearby, bright stars for faint companions, which is of great importance for
studies on hot exozodiacal dust and target vetting for future exo-Earth
imaging missions such as NASA’s Habitable Worlds Observatory. A further
application is aperture synthesis of extended sources such as active galactic
nuclei or evolved stars.
\end{abstract}


\keywords{
    aperture masking,
    interferometry,
    integral field spectrograph,
    Large Binocular Telescope Interferometer (LBTI),
    Arizona Lenslets for Exoplanet Spectroscopy (ALES),
    LMIRcam,
    infrared,
    high-contrast methods
}


\section{INTRODUCTION}
\label{sect_intro}

Non-redundant aperture-masking interferometry (NRM), also called sparse
aperture-masking interferometry, is a technique to modify a single telescope
to act as an interferometric array of multiple telescopes\cite{haniff:1987,
readhead:1988, nakajima:1989, tuthill:2012}. This is achieved by inserting an
opaque mask with holes in the pupil plane of the telescope. The subapertures
formed by the holes produce a distinct point spread function—in this context
also called interferogram—that can be interpreted via Fourier analysis. The
holes are arranged so that each vector connecting the centers of two holes, the
baseline, is unique, which decreases the effects of atmospheric turbulence.
Moreover, this non-redundancy of the baselines allows for measuring
high-quality closure phases\cite{baldwin:1986}. This quantity, originating in
radio interferometry\cite{jennison:1958, rogers:1974, pearson:readhead:1984},
is sensitive to deviations from point-symmetry in the astrophysical source and
is computed by summing the phases of three baselines forming a closed triangle,
which cancels out atmospheric phase contributions.

NRM allows probing the source at and below the diffraction limit of the full
telescope aperture while the majority of the collected light is blocked by
the mask (which can be mitigated\cite{doelman:2018, doelman:2021, taras:2025,
piroscia:2026}). The technique originally required short exposures, similar in
length to the atmospheric coherence time, to record the interference pattern,
which starkly limited the applicability to very bright targets. However, the
rise of high-performance adaptive optics now allows for integration times much
longer than the coherence time of the measured light\cite{tuthill:2006}.
Today, several major ground- and space-based facilities are equipped with NRM
capabilities, such as the instruments SPHERE\cite{cheetham:2016, beuzit:2019}
and ERIS\cite{davies:2023} at the Very Large Telescope,
GPI\cite{macintosh:2006, macintosh:2014a, macintosh:2014b, greenbaum:2019} at
the Gemini Observatory (currently being upgraded to GPI 2.0\cite{chilcote:2020,
peng:2023, wolff:2026}), SCExAO\cite{guyon:2010} at the Subaru Telescope,
SCALES\cite{lach:2022, skemer:2022, sallum:2023b, stelter:2024, lach:2026} and
NIRC2 at the W. M. Keck Observatory\cite{tuthill:2000}, the Large Binocular
Telescope Interferometer (LBTI)\cite{hinz:2012, hinz:2014, hinz:2016,
ertel:2020b} of the Large Binocular Telescope (LBT)\cite{hill:2014, wagner:2014,
rothberg:2016, shields:2024}, MagAO-X\cite{males:2018, close:2018, males:2022,
males:2024} at the Magellan Clay Telescope, or NIRISS\cite{doyon:2012,
doyon:2023, artigau:2014, sivaramakrishnan:2023, sallum:2024} on-board the
James Webb Space Telescope\cite{gardner:2006}.

Spatial information about the astrophysical source retrieved by NRM depends on
the number of spatial frequencies ($u$, $v$) sampled in the Fourier plane. A
powerful method for improving the Fourier plane sampling is to spectrally
disperse the measurements. As the spatial frequencies are defined by the ratio
of baseline and observing wavelength $\lambda$, spectrally dispersing the
interferogram results in a more finely sampled $u$-$v$-plane. While exploiting
sky rotation with several pointings fills the $u$-$v$-plane azimuthally,
spectral dispersion fills it radially (see Sect.~\ref{sect_data_reduction} and
Fig.~\ref{fig_uv_coverage}). In addition, spectral dispersion reduces the
effect of spectral bandwidth smearing. First successes of this technique were
achieved with MAPPIT\cite{robertson:1991, bedding:1993} at the Anglo-Australian
Telescope\cite{marson:1992, bedding:1993} by combining a dispersive element
with a special NRM mask, which featured an assembly of square-shaped holes
aligned perpendicular to the direction of spectral dispersion. Two-dimensional
probing of the astrophysical source required multiple pointings, exploiting sky
rotation. This technique was later used again at the W. M. Keck observatory to
measure spectrally dispersed NRM across the $J$ through $L$ bands up to
${\sim} \qty{4}{\um}$\cite{woodruff:2009}.

The rise of integral field spectrographs (IFSs) that enable spectrally
dispersed imaging allowed to push spectrally dispersed NRM further. IFSs allow
for using classical NRM masks with a two-dimensional hole layout, delivering
richer spatial information per pointing than the previously required linear
hole assemblies. This was pioneered at the Hale Telescope of Palomar
Observatory\cite{zimmermann:2012} and is nowadays a capability of SPHERE and
GPI 2.0. However, combining an IFS with NRM was never achieved at wavelengths
longer than $K$ band (${\sim} \qty{2}{\um})$. A significant challenge at longer
wavelengths is that the thermal background increases drastically with
increasing wavelength. After the initial success at the W. M. Keck Observatory
to measure spectrally dispersed NRM in the $L$ band (not with an IFS, see above)
that targeted extremely bright Mira variable stars with a $K_\textrm{s}$-band
magnitude of $< \num{-2.21}$\cite{cutri:2003}, no further observations have
been reported.

In searches for binary stars, performing NRM in the thermal infrared $L$ and
$M$ bands has the advantage of increasing the flux ratio between the typically
redder, fainter companion star to the brighter primary star. Our particular
interest in this technique comes from the need to perform companion searches
around stars hosting hot exozodiacal dust\cite{absil:2006, absil:2013,
ertel:2014, kral:2017, nunez:2017, absil:2021, ertel:2025} after previous
methods to exclude companions based on data from long-baseline interferometry
were shown to be flawed\cite{tsishchankava:2026} and a companion was recently
detected in a frequently observed system\cite{stuber:2026a}. Known host stars
of hot exozodiacal dust are typically G- to A-type stars, which causes possible
companion stars to be typically K- to M-type dwarf stars, or brown dwarf stars.
While a thorough understanding of the hot exozodiacal dust phenomenon remains
elusive\cite{pearce:2022b, ertel:2025}, it jeopardizes attempts to image
exo-Earths\cite{ertel:2025, stapelfeldt:mamajek:2025}, for instance with the
future Habitable Worlds Observatory (HWO)\cite{feinberg:2026} or Large
Interferometer for Exoplanets (LIFE)\cite{quanz:2022}, and an improved
characterization of hot-dust hosting systems is wanted. Furthermore, NRM
delivers a wealth of spatial information that facilitates image reconstruction
using aperture synthesis; deeply red objects such as active galactic nuclei or
stars undergoing late-stage evolution are natural targets.

In this proceeding, we report on the first execution of spectrally dispersed
NRM observations in the $L$ band using an IFS and a mask with two-dimensional
hole layout. To achieve this, we combined the IFS Arizona Lenslets for
Exoplanet Spectroscopy (ALES)\cite{skemer:2015, hinz:2018, skemer:2018,
stone:2018, stone:2022} of the LBTI with NRM. While commissioning was performed
in the $L$ band, our setup is suitable for $M$-band observations by changing
the spectral mode of ALES. We explain the instrumental setup in
Sect.~\ref{sect_inst_setup}, the commissioning observations of a known binary
star in Sect.~\ref{sect_observations}, and the data reduction in
Sect.~\ref{sect_data_reduction}. In Sect.~\ref{sect_companion_search}, we show
the analysis of the binary star observations with the recovery of the fainter
companion's position and derivation of detection limits. We discuss future
prospects for improving the new observing mode in
Sect.~\ref{sect_future_prospects} before we close with a summary in
Sect.~\ref{sect_summary}.

\begin{table}
    \centering
    \caption{
        Hole positions of the LBTI 12-hole mask as projected on the telescope's
        primary aperture from the design specification. This work uses only the
        left mirror (SX).
    }
    \vspace{0.1in}
    \begin{tabular}{cSS|cSS}
        \toprule
        \multicolumn{3}{c|}{Left mirror (SX)} &
        \multicolumn{3}{c}{Right mirror (DX)} \\
        Hole Id. & x~/\unit{\meter} & y~/\unit{\meter} & Hole Id. &
        x~/\unit{\meter} & y~/\unit{\meter} \\
        \midrule
        H1  & -4.21950  & -1.71358 & H7  & 5.62600  &  3.15868 \\
        H2  & -4.92275  & -0.49552 & H8  & 4.92275  &  1.94061 \\
        H3  & -4.92275  &  1.94061 & H9  & 4.21950  & -1.71358 \\
        H4  & -9.14225  & -2.93164 & H10 & 9.14225  & -2.93164 \\
        H5  & -10.54870 & -0.49552 & H11 & 10.54870 & -0.49552 \\
        H6  & -7.03250  &  3.15868 & H12 &  7.03250 &  3.15868 \\
        \bottomrule
    \end{tabular}    
    \label{table_hole_positions}
\end{table}


\section{INSTRUMENTAL SETUP}
\label{sect_inst_setup}

Non-redundant aperture masks (hereafter: aperture masks) and ALES are both
integrated into the LBTI's $L$/$M$-band Infrared Camera
(LMIRcam)\cite{wilson:2008, leisenring:2010, skrutskie:2010, kuzmenko:2012,
leisenring:2012}. Our instrument setup is that of a standard ALES observation
with an aperture mask inserted into the light path. Here we present the most
relevant parts of the setup and their principle of operation; a comprehensive
schematic can be found in Ref.~\citenum{stone:2018}, Fig.~1, and an
illustration of ALES' working principle in Ref.~\citenum{skemer:2018}, Fig.~1.

Light is collected by one of the two \qty{8.4}{\meter} LBT primary mirrors
while the wavefront is corrected by the respective adaptive secondary
mirror\cite{gallieni:2000, gallieni:2003, riccardi:2003, riccardi:2010,
briguglio:2012, guerra:2013, christou:2014, brusa:2020, zhang:2020, brusa:2024,
zhang:2024} that is part of the LBT's adaptive optics system\cite{bailey:2014,
pinna:2016, pinna:2021, pinna:2023} (based on the earlier
FLAO\cite{esposito:2003a, esposito:2003b, quiros_pacheco:2010, esposito:2010a,
esposito:2010b}). After three warm reflections, the light enters the cryogenic
LBTI and is directed to LMIRcam. The beam passes through the aperture mask
located in the first LMIRcam filter wheel (pupil plane) and gets magnified by a
Keplerian magnifier (currently a refractive one as in the original ALES
setup\cite{skemer:2015}) for sufficient spatial sampling by the
\qty{500}{\um}-pitch lenslet array\cite{skemer:2018}. This lenslet array (focal
plane) then focuses the area associated with each lens, producing physically
separated spatial pixels (spaxels). Using a prism, these spaxels are spectrally
dispersed, producing physically separated spectra on the detector.

We used the left (SX) LBT mirror and the twelve-hole aperture mask (see
Ref.~\citenum{leisenring:2012}, Fig.~6, top), which has six holes for each
mirror. The hole positions projected on the telescope's primary aperture
($x$, $y$) are listed in Table~\ref{table_hole_positions}; the hole diameter is
\qty{0.784}{\meter}. The numbers reflect the positions according to the design,
but the number of digits does not signify manufacturing precision. In our
setup, using only the SX mirror results in an effective six-hole mask with
\num{15} baselines ranging in length from ${\approx} \qty{1.41}{\m}$ to
${\approx} \qty{6.45}{\m}$ and \num{20} closure phase triplets of which
\num{10} are non-redundant\cite{readhead:1988, monnier:2003b}. We chose the
$L$-band mode for ALES, spectrally dispersing the light approximately between
\qty{2.8}{\um} and \qty{4.3}{\um} with a spectral resolving power of
$R \sim \num{40}$\cite{skemer:2018}.

\begin{figure}
    \centering
    \includegraphics[width=0.7\linewidth]{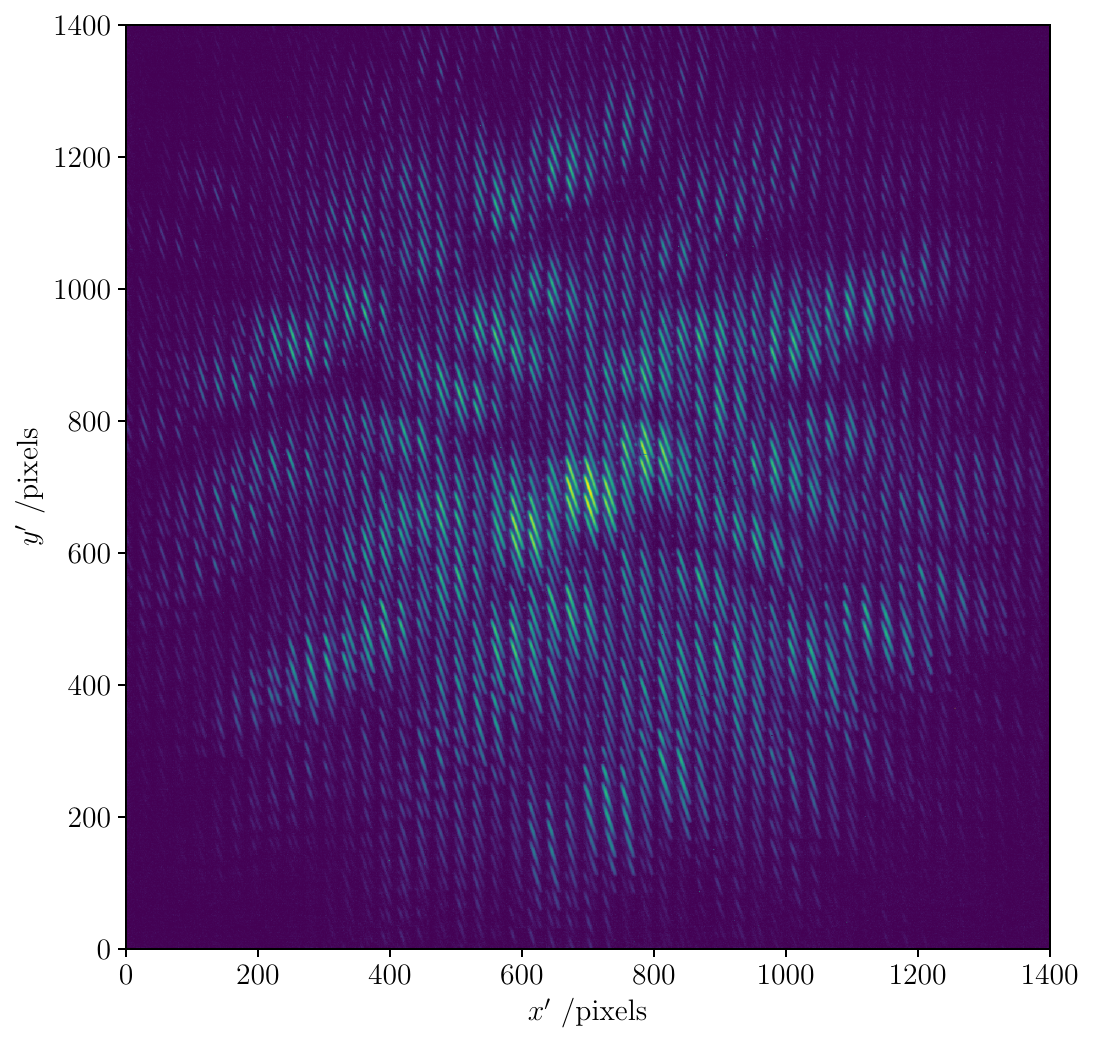}
    \caption{
        Section of LMIRcam detector frame ($x'$, $y'$)
        using ALES and the aperture mask on the star HD 75156 (see
        Sect.~\ref{sect_observations}). The background was subtracted and bad
        pixels were corrected. The colormap uses an inverse hyperbolic sine
        function to optimize visual clarity.
    }
    \label{fig_detector_image}
\end{figure}

A detector frame for the target star HD 75156 (see
Sect.~\ref{sect_observations}) with background subtracted and bad pixels
corrected is shown in Fig.~\ref{fig_detector_image}. On small scales, the
individual spectra are visible as streaks, while on large scales, the
interferogram dominates.

\section{OBSERVATIONS}
\label{sect_observations}

We observed the known binary system $\epsilon$ Hydrae ($\epsilon$ Hya,
HD 74874, HIP 43109) on 8 November 2025. The G1III and A8V
pair\cite{stephenson:sanwal:1969} is locked in an orbit with a
${\approx} \qty{15}{yr}$ period and a semi-major axis of
${\sim} \qty{250}{mas}$\cite{anguita_aguero:2023}. The projected separation of
the binary was $\qty{202(5)}{mas}$ in January 2023\cite{guerrero:2025} and its
joint $K_\mathrm{s}$-band magnitude is ${\approx} \num{1.3}$.\cite{cutri:2003}
As calibrator star, we selected HD 75156 (HIP 43251, V* FX Cancri) from the
MDFC catalog\cite{cruzalebes:2019}. We observed the science target (SCI) and
calibrator star (CAL) in the sequence CAL-SCI-CAL-SCI-CAL-SCI-CAL-SCI-CAL.

For each of the four SCI and first four CAL blocks, we nodded two times between
target and sky to record \num{25} frames of $\qty{4.4258}{\second}$ integration
time per nod position for a total of \num{50} frames with a total integration
time on target of ${\approx} \qty{221}{\second}$ per block. For the last CAL
block, only one nod cycle was completed for  ${\approx} \qty{111}{\second}$ on
target. The total integration time on the science target across the whole
sequence was ${\approx} \qty{885}{\second}$, covering parallactic angles from
\qty{-34.7}{\degree} to \qty{5.6}{\degree}. The execution time for the entire
sequence was ${\sim} \qty{2}{\hour}$.

Our observations were impaired by water absorption and thus we restrict all
following analysis to wavelengths between \qty{3.5}{\um} and \qty{4}{\um} were
the signal-to-noise ratio is highest and the scatter of the extracted
interferometric observables (see Sect.~\ref{sect_data_reduction}) small.
However, with more favorable conditions we expect a broader spectral range to
be exploitable, approximately between \qty{3.1}{\um} and \qty{4.1}{\um}.

In addition, prior to executing the NRM sequence, we recorded images using
LMIRcam without ALES and the aperture mask to verify the results obtained with
the new spectrally dispersed NRM mode, in particular the retrieved position of
the fainter binary component relative to the primary. We used a neutral density
filter with \qty{10}{\percent} transmission and nodded two times between the
left and right side of the detector, recording \num{50} frames per nod position
on target and background simultaneously. With \num{200} frames on target with
an integration time of $\qty{0.014}{\second}$ each, this yields a total
integration time on target of $\qty{2.8}{\second}$.


\section{DATA REDUCTION}
\label{sect_data_reduction}

\begin{figure}
    \centering
    \includegraphics[width=1\linewidth]{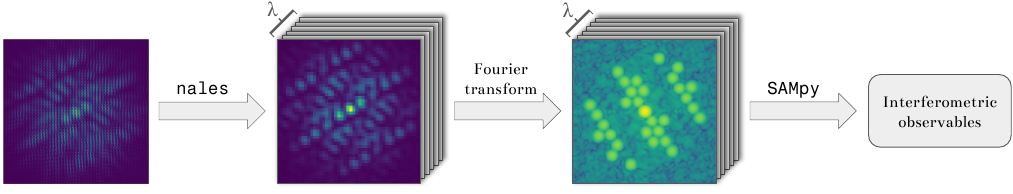}
    \caption{
        Data reduction sketch. From the detector image with background
        subtracted and bad pixels corrected (left), three-dimensional spectral
        cubes are extracted using \texttt{nales}, containing for each
        wavelength the interferogram (second from left). Of each interferogram,
        the Fourier transform is computed, yielding for each wavelength the
        complex visibility (second from right, here visualized by the power
        spectrum). These are used to determine the mask rotation angle (see
        Sect.~\ref{sect_data_reduction}). Lastly, from the complex
        visibilities, spectrally dispersed interferometric observables—closure
        phases, squared visibilities, amplitudes and phases of the complex
        visibility—are extracted using \texttt{SAMpy} (right).
        }
    \label{fig_data_reduction_sketch}
\end{figure}

The major steps of the data reduction are illustrated in
Fig.~\ref{fig_data_reduction_sketch}. The first one is to produce spectral
cubes (two-dimensional images with wavelength as third dimension) from the raw
LMIRcam frames using standard ALES procedures, for which we used the Python
package \texttt{nales}\cite{stone:2024}. For each wavelength and across frames,
we center the interferograms by first determining the center of each frame by
fitting a two-dimensional Gaussian function to the center fringe, then shift
all frames to a common center using Fourier methods. Then, groups of five
frames are stacked by replacing each pixel with the median pixel value from all
frames. Subsequently, by computing the Fourier transform of the interferograms,
the complex visibility and the power spectrum (proportional to the squared
absolute of the complex visibility) is computed for each wavelength and
(stacked) frame.

To determine the baselines and thus spatial frequencies, the orientation angle
of the aperture mask in the filter wheel needs to be calibrated. We use the
hole positions from Table~\ref{table_hole_positions} to compute the position of
the peaks in the interferogram's power spectrum (see
Fig.~\ref{fig_data_reduction_sketch}, second from right). Then, we rotate the
hole positions to match the predicted peak positions to the measured peaks of
a calibrator observation. The alignment routine first performs a coarse angular
grid search to find the approximate mask rotation angle. Then, it minimizes the
squared sum of the differences of the predicted peak positions and the center
of a two-dimensional Gaussian function fitted to the actual peaks in the power
spectrum of the calibrator observation. This is performed across wavelengths
and frames, with the mean and standard deviation of all measurements used as
the mask rotation angle and its uncertainty, respectively. Using all CAL
blocks, we estimate a counter-clockwise mask rotation angle of
\qty{4.256(0.018)}{\degree}. This calibration should be performed with each
data set as the filter wheel position is not fully repeatable and changes with
pupil alignment.

\begin{figure}
    \centering
    \includegraphics[width=1\linewidth]{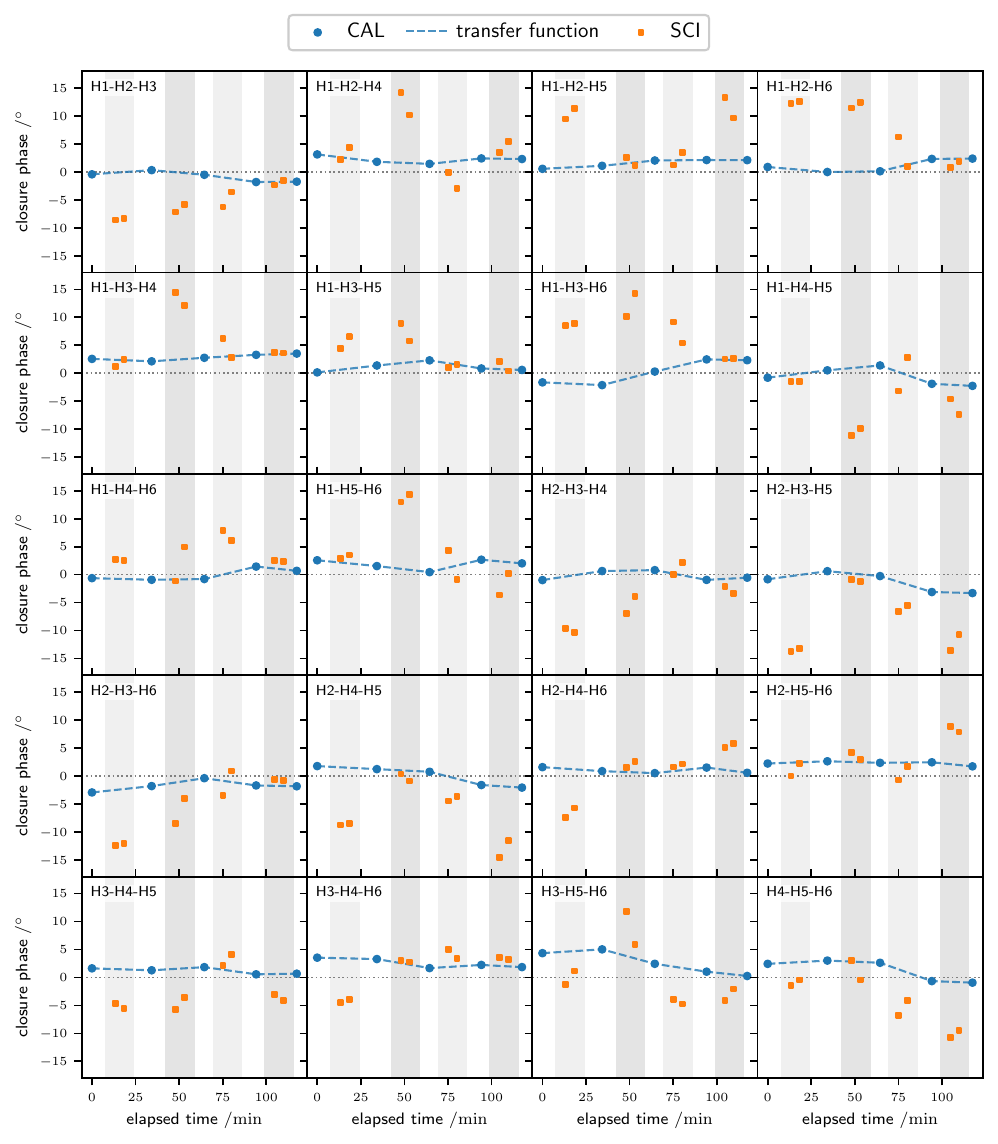}
    \caption{
        Closure phases of our observing sequence for all \num{20} triplets,
        averaged over wavelengths. Phases of calibrator observations (CAL) are
        shown with blue circles; the dashed blue lines that linearly connect
        the calibrator measurements estimate the instrumental transfer function
        during the observing sequence. Phases of science target observations
        (SCI) are shown with orange squares. While the calibrator phases are
        close to zero, the science target phases show a stronger signal subject
        to temporal change due to the binarity of the source.
    }  
    \label{fig_transfer_function}
\end{figure}

The extraction of interferometric observables is performed by the numeric tool
\texttt{SAMpy}\cite{sallum:eisner:2017, sallum:2022} in the implementation of
Ref.~\citenum{stone:2026}. We extract closure phases, squared visibilities, and
complex visibilities (consisting of visibility amplitudes and visibility
phases). For our bright sources, data reduction with or without frame stacking
yields compatible observables. In the following, we will focus on the closure
phases. For the calibrator blocks, we extract one set of observables from all
\num{50} frames (\num{10} stacked) per block. For the science blocks, we
extract two sets from \num{25} frames (\num{5} stacked) each to reduce smearing
of the companion signal due to sky rotation. The respective uncertainties are
computed as the standard deviation among frames withing the same block. This
frame-to-frame scatter is on the order of ${\sim} \qty{0.1}{\degree}$.
Figure~\ref{fig_transfer_function} shows the closure phases for each triplet,
averaged across wavelengths. The closure phases of the calibrator observations
are close to zero with the absolute of phases being always smaller than
\qty{5}{\degree}; these measurements provide an estimate of the instrumental
transfer function—the closure phase signal caused by the instrument. On the
other hand, the observations of the binary star show more extreme closure
phases that change over time, thus with sky rotation.

\begin{figure}
    \centering
    \includegraphics[width=0.6\linewidth]{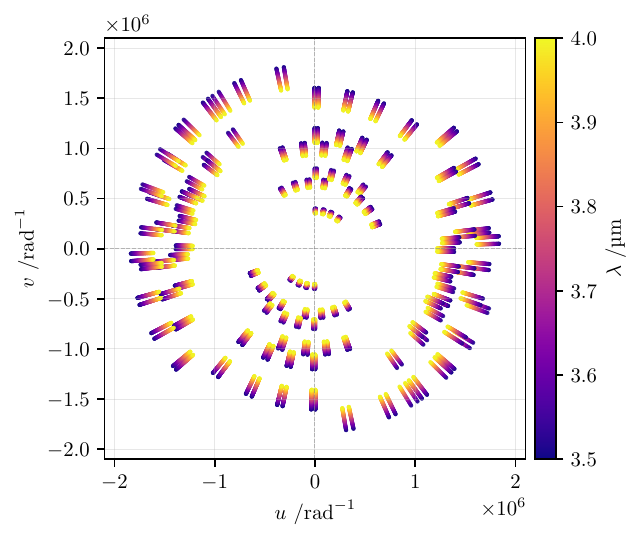}
    \caption{
        Spatial frequencies ($u$, $v$) sampled by our observations, restricted
        to wavelengths $\lambda$ (color-coded) with high-signal-to-noise ratio.
    }
    \label{fig_uv_coverage}
\end{figure}

Each set of science observables is calibrated against the linear interpolation
in time between the observables of the bracketing calibrator observations. For
squared visibilities and visibility amplitudes, calibration is a division of
the calibrator observable; for closure and visibility phases, it is a
subtraction. Propagating the frame-to-frame scatter through the calibration
procedure yields closure phase uncertainties of up to
${\sim} \qty{0.2}{\degree}$. This is small compared to the standard deviation
among closure phases from all calibrator blocks that is on the order of
${\sim} \qty{2}{\degree}$ (see Fig.~\ref{fig_transfer_function}). Therefore,
we adopt this scatter across the entire sequence as the uncertainty of the
calibrated closure phases.

Data reduction results in eight sets of interferometric observables of
$\epsilon$~Hya. These are paired with the baselines derived directly from the
rotated positions of the mask holes; see Fig.~\ref{fig_uv_coverage} for the
Fourier plane coverage of the observing sequence.

All reduction steps after cube extraction with \texttt{nales}, including saving
the raw or calibrated observables in the OIFITS2 format\cite{duvert:2017}, are
handled with the new Python package \texttt{ales-nrm}\cite{stuber:2026b}. While
it currently uses \texttt{SAMpy} as the backend for extraction, its design
allows for the easy integration of other backends or a native implementation.


\section{COMPANION SEARCH}
\label{sect_companion_search}

\begin{figure}
    \centering
    \includegraphics[width=0.49\linewidth, valign=t]{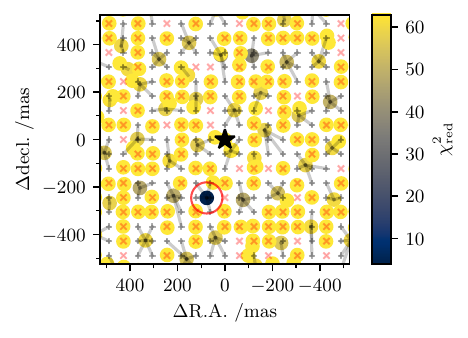}
    \includegraphics[width=0.49\linewidth, valign=t]{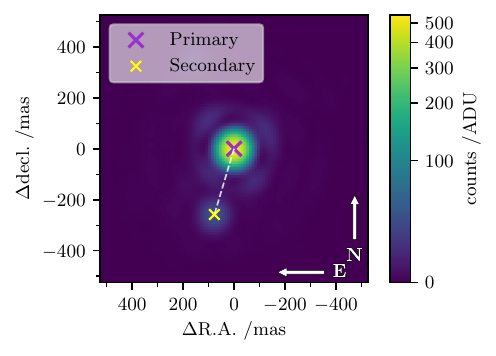}
    \caption{
        Left: Result from the \texttt{PMOIRED} grid search in $\Delta$decl. and
        $\Delta$R.A. for the companion around the primary star (black) of
        $\epsilon$~Hya in the coordinate center. Grey lines connect initial
        locations of a single fit (grey plus signs) with the best-fit location.
        Red crosses denote initial positions for which the respective fit did
        not converge. Filled circles denote the position for the last fit
        iteration for both converged and not converged fits; the circle color
        indicates the value of the reduced $\chi^2$, $\chi^2_\textrm{red}$. The
        companion is found at the location of the smallest value of
        $\chi^2_\textrm{red} \approx \num{4}$ (dark filled circle) that is
        further indicated by the red circle.
        Right: LMIRcam image of the $\epsilon$ Hya binary star with the
        colorbar scaled with an inverse hyperbolic sine function.
    }
    \label{fig_grid_search_lmircam_image}
\end{figure}

To verify our observing setup, we constrain the position of the secondary
component of $\epsilon$~Hya using our spectrally dispersed NRM observations and
compare the results to that of LMIRcam imaging. Fitting only the closure
phases, we use the numerical tool \texttt{PMOIRED}\cite{merand:2022,
merand:2024} to perform a grid search over declination and right ascension
relative to the primary star ($\Delta$decl., $\Delta$R.A.) using the
\texttt{CANDID} algorithm\cite{gallenne:2015} and clearly detect the companion
(see Fig.~\ref{fig_grid_search_lmircam_image}, left). Using
bootstrapping\cite{efron:1979, efron:1982} with \num{2000} samples and each
sample fit being initialized with the global solution from the grid search, we
sample the closure phase triplets to determine the final companion positions
and uncertainties. The bootstrapping algorithm treats spectrally dispersed data
from one triplet as a group and samples from all triplets. Thereby, data
belonging to a certain triplet is treated fully
correlated\cite{ertel:2014, stuber:2026a}. We retrieve a binary separation of
$\rho_\textrm{NRM} = \qty{257.3(0.7)}{mas}$ and a position angle of the
companion (\qty{0}{\degree} is up, increasing counterclockwise) of
$\theta_\textrm{NRM} = \qty{162.7(0.1)}{\degree}$. The uncertainty of the
position angle includes the small uncertainty of the mask orientation
(see Sect.~\ref{sect_data_reduction}).

For comparison, we retrieve the companion position from the LMIRcam image. To
determine the positions of the primary and secondary component, we fit
two-dimensional Gaussian functions to image cutouts centered on the pixel with
the highest local flux and a radius of four pixels, respectively. Based on the
positions, we compute the position angle of the companion
$\theta_\textrm{img}$, and the binary separation $\rho_\textrm{img}$ by using
the LMIRcam plate scale of \qty{10.651(0.038)}{mas}. To compute the
uncertainties, we propagate those from the position fits and combine them in
quadrature to that of the plate scale (for $\rho_\textrm{img}$) and to that of
the determination of true north (\qty{0.236}{\degree}, for
$\theta_\textrm{img}$). We yield a separation of
$\rho_\textrm{img} = \qty{269(1)}{mas}$ and a position angle of
$\theta_\textrm{img} = \qty{163.3(0.2)}{\degree}$.

Using our spectrally dispersed NRM observations, we find the companion at
approximately the same position as shown by the LMIRcam imaging and confirm the
functionality of our experimental setup. While the determined position angles
agree, however, there is a separation discrepancy of ${\approx} \qty{12}{mas}$.
The mismatch in separation could originate in a biased determination of the
spatial frequencies, either via biased hole positions or uncertainties of the
wavelength calibration (see Sect.~\ref{subsect_future_prospects_execution}).

\begin{figure}
    \centering
    \includegraphics[width=0.7\linewidth]{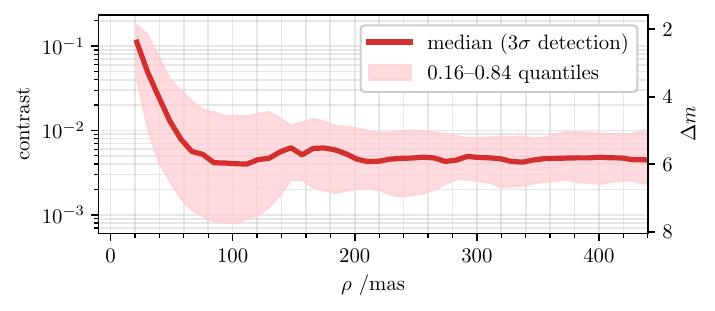}
    \caption{
        Contrast limit for the detection of a third stellar component in the
        system of $\epsilon$~Hya. The curve is computed as the median contrast
        within radial bins of a randomized grid of injected components. The
        pink region covers the region between the \num{0.16} and \num{0.84}
        quantiles within each radial bin.
    }
    \label{fig_contrast_curve}
\end{figure}

To derive the detection limit for a possible third component, we use the method
of Ref.~\citenum{absil:2011} that is implemented in \texttt{PMOIRED}. First, we
assume the found binary model to be correct. Then, a third stellar component is
injected in a randomized pattern and its flux is increased as far as necessary
to yield a $3\sigma$ detection (with $\sigma$ being the standard deviation of a
Gaussian function), that is with a false-detection probability of
\qty{0.27}{\percent}. Lastly, we compute radial profiles of the median contrast
between injected component and primary star within radial bins of the
randomized pattern (see Fig.~\ref{fig_contrast_curve}). We reach a contrast of
${\sim} \num{5e-3}$ for distances to the primary star of
${\gtrsim} \qty{80}{mas}$.


\section{FUTURE PROSPECTS}
\label{sect_future_prospects}

Several steps can be taken to improve the performance of the new observing mode
and advance its capabilities.


\subsection{Execution}
\label{subsect_future_prospects_execution}

One can record auxiliary images using ALES' fixed-frequency dot
grids\cite{skemer:2015}. Using them to correct distortion in the interferograms
might improve the sensitivity and precision of the complex visibilities. It
would have no effect on the computed baselines and spatial frequencies as these
are directly derived from the mask geometry.

In addition, one can image the aperture mask in a pupil plane, effectively
imaging the circular holes. This would allow for an independent confirmation of
the determined mask rotation angle and would allow to investigate the position
of the mask holes at time of observation. Thereby, bending of the mask due to
thermal effects during cryostat cool-down and warm-up, or hands-on work on the
system could be spotted. This might lead to a correction of the mask hole
positions (Table~\ref{table_hole_positions}) and thus reduced bias of the
spatial frequencies. However, any apparent distortion of the mask in a pupil
plane image would be mixed with distortion due to the LMIRcam optical train.
If the hole positions and thus baselines are robustly determined, remaining
mismatches between true and determined binary star separations would originate
in bias of the wavelength calibration. 


\subsection{Spectral Windows}

While our commissioning observations employed the established $L$-band mode of
ALES, further not fully commissioned modes offer spectral windows across the
$K$ through $M$ bands (see Ref.~\citenum{skemer:2018}, Table 2). Especially the
$L$/$M$-band mode ($\sim$\num{3}–\qty{5}{\um} with $R \sim \num{20}$) would
provide access to even longer wavelengths that were never probed with
spectrally dispersed NRM observations. Observing at longer wavelengths would
further lower the contrast requirements for detecting sources that are cooler
compared to the primary star such as low-mass companions or dust distributions.
However, while nothing prevents this mode from execution, its sensitivity needs
to be tested; at longer wavelengths, sources are typically fainter while the
intensity of the thermal background increases. Furthermore, narrow-band filters
for calibration are currently only available up to ${\sim} \qty{4.1}{\um}$,
impairing the achievable precision of the wavelength calibration in the $M$
band.


\subsection{Software}

Efforts to improve the ALES data reduction software \texttt{nales} are
ongoing\cite{li:2026} and future reductions might use up-the-ramp
fitting\cite{isbell:2026}, which would improve the sensitivity and thus
performance for faint sources. The \texttt{ales-nrm} package is in active
development.


\subsection{Double-sided Fizeau Interferometric Imaging}

While we used only one mirror of the LBT, NRM with ALES can be executed using
both mirrors together, performing Fizeau interferometric imaging with baseline
lengths up to ${\approx} \qty{21.1}{\meter}$. However, a magnification
surpassing the currently used eightfold one is required to achieve a
Nyquist-sampled interferogram when using ALES. Suitable optical magnifiers are
available, but they need to be aligned within the optical train; this is
foreseen to be performed in Summer 2027. Furthermore, for higher magnifications
the core of the point spread function (see Fig.~\ref{fig_detector_image}) will
be larger than the detector and thus truncated. If the recorded part would not
allow the computation of complex visibilities of sufficient quality, a
technical solution needs to be found.

\section{SUMMARY}
\label{sect_summary}

The LBTI is now able to perform spectrally dispersed non-redundant
aperture-masking interferometry in the thermal infrared. We have commissioned
this new mode by observing a known binary star and recovering the fainter
secondary component using fully reduced and calibrated closure phases. We
estimate our contrast detection limit to be ${\sim} \num{5e-3}$ for separations
to the primary star of ${\gtrsim} \qty{80}{mas}$. While the new mode is fully
operational in $L$ band and using one \qty{8.4}{\meter} LBT mirror at a given
time, efforts are ongoing to extend the spectral coverage into the $K$ and $M$
bands as well as enabling double-sided operations, unlocking the full
interferometric capabilities of the LBTI for spectrally dispersed non-redundant
aperture-masking interferometry.


\acknowledgments

The authors T.A.S., J.P.S., and S.E. acknowledge support by the National
Aeronautics and Space Administration (NASA) through grant 80NSSC23K1473 while
T.D.P. acknowledges support of the Research Foundation - Flanders (FWO) through
grant 11P6I24N.

All authors acknowledge the use of the Large Binocular Telescope Interferometer
(LBTI) and the support from the LBTI team, specifically from Alexander Becker.
The LBT is an international collaboration among institutions in the United
States and Europe. At the time data were acquired for this research, LBT
Corporation Members were The University of Arizona on behalf of the Arizona
Board of Regents; Istituto Nazionale di Astrofisica, Italy; LBT
Beteiligungsgesellschaft, Germany, representing the Max-Planck Society, the
Leibniz Institute for Astrophysics Potsdam, and Heidelberg University; and The
Ohio State University, representing The Ohio State University, University of
Notre Dame, University of Minnesota, and University of Virginia. This research
used the facilities of the Italian Center for Astronomical Archives (IA2)
operated by INAF at the Astronomical Observatory of Trieste. Observations have
benefited from the use of ALTA Center
(\href{https://alta.arcetri.inaf.it/}{alta.arcetri.inaf.it}) forecasts
performed with the Astro-Meso-Nh model. Initialization data of the ALTA
automatic forecast system come from the General Circulation Model (HRES) of
the European Centre for Medium Range Weather Forecasts.

The large language model \texttt{Claude Opus 4.6} has been used for software
development on \texttt{ales-nrm} as well as analysis and plotting routines;
access was granted via University of Arizona Generative AI. Furthermore, this
research has made use of
\texttt{astropy}\cite{astropy_collaboration:2013,
astropy_collaboration:2018, astropy_collaboration:2022},
\texttt{Ipython}\cite{perez:granger:2007},
\texttt{Jupyter} notebooks\cite{kluyver:2016},
\texttt{Matplotlib}\cite{hunter:2007},
\texttt{NumPy}\cite{harris:2020},
\texttt{OIFITS}\cite{duvert:2017},
\texttt{PMOIRED}\cite{merand:2022, merand:2024},
\texttt{Python} (\href{https://www.python.org/}{https://www.python.org/}),
\texttt{SIMBAD} Astronomical Database\cite{wenger:2000},
\texttt{VizieR}\cite{ochsenbein:2000},

the Jean-Marie Mariotti Center (JMMC) service \texttt{OIFits Explorer}, the
Washington Double Star Catalog maintained at the U.S. Naval Observatory, and
data products from the Two Micron All Sky Survey, which is a joint project of
the University of Massachusetts and the Infrared Processing and Analysis
Center/California Institute of Technology, funded by NASA and the National
Science Foundation.

\bibliography{bibliography}
\bibliographystyle{spiebib}

\end{document}